\documentclass[11pt,a4paper]{article}
\pdfoutput=1

\usepackage{jheppub}
\usepackage{rotating}

\usepackage{wrapfig}

\usepackage{amsfonts}
\usepackage{amsmath}
\usepackage{slashed}
\usepackage{amssymb}
\usepackage{latexsym}
\usepackage{multirow}
\usepackage{hyperref}
\usepackage{enumitem}
\hypersetup{colorlinks=true,citecolor=red,linkcolor=magenta,urlcolor=blue}
\usepackage{relsize}
\usepackage{booktabs}
\usepackage{subfigure}
\usepackage{cleveref}
\usepackage{fancyhdr}
\usepackage{float}
\usepackage{graphicx}
\usepackage{microtype}
\usepackage{tabularx}
\usepackage{xcolor}
\usepackage{siunitx,adjustbox,bm}
\usepackage[bottom]{footmisc}

\usepackage{physics}

\renewcommand{\arraystretch}{1.3}

\newcolumntype{C}{>{$}c<{$}}
\newcolumntype{L}{>{$}l<{$}}
\newcolumntype{R}{>{$}r<{$}}

\title{Two Higgs doublets, Effective Interactions and a Strong First-Order Electroweak Phase Transition}
\author[a,b]{Anisha,}
\author[c]{Lisa Biermann,}  
\author[b]{Christoph Englert,} 
\author[c]{and Margarete M\"uhlleitner}
\affiliation[a]{Department of Physics, Indian Institute of Technology, Kanpur-208016, India}
\affiliation[b]{School of Physics \& Astronomy, University of Glasgow, Glasgow G12 8QQ, United Kingdom}
\affiliation[c]{Institute for Theoretical Physics, Karlsruhe Institute of Technology, 76128 Karlsruhe, Germany}
\emailAdd{anisha@iitk.ac.in}
\emailAdd{lisa.biermann@kit.edu}
\emailAdd{christoph.englert@glasgow.ac.uk}
\emailAdd{margarete.muehlleitner@kit.edu}

\abstract{It is well-known that type II two Higgs doublet models (2HDMs) can struggle to facilitate a strong first-order electroweak phase transition in the early universe whilst remaining theoretically appealing scenarios for many reasons. We analyse this apparent shortfall 
from the perspective of additional new physics. Starting from a
consistent dimension-6 effective field theory Higgs potential
extension, we identify the Higgs potential
  extensions that provide the necessary additional contributions required to achieve a strong first-order electroweak phase transition and trace their phenomenological implications for the Large Hadron Collider. 
In passing, we critically assess the reliability of the dimension-6
  approximation depending on the expected 2HDM phenomenology. In
  particular, we focus on the role of Higgs pair production (resonant and non-resonant) and interference effects expected in top final states, which are the prime candidates of 2HDM exotics discoveries.}

\begin{document}
\maketitle
\section{Introduction}
\label{sec:intro}
The null results of searches for new physics beyond the Standard Model
(BSM) chiefly performed at the Large Hadron Collider (LHC) have left
particle physics in a delicate status quo: The standard paradigms that
have shaped BSM model building for the past decades stand challenged,
and the role of the TeV scale in nature alongside its microscopic
origin are profoundly unclear. This observation is accompanied by
insurmountable evidence that new physics is required to reconcile
physics at the smallest distances with astrophysical and cosmological
observations. 
The Sakharov
criteria~\cite{Sakharov:1967dj} provide a strong motivation to
incorporate additional sources for CP violation and dynamics
responsible for a strong first-order electroweak phase transition to our particle physics picture for efficient baryogenesis. There are various ways to achieve the latter which venture away from minimal SM extensions~\cite{Affleck:1984fy,Espinosa:2011ax,Niemi:2020hto}. Nonetheless electroweak baryogenesis remains an attractive avenue and the potential implications for TeV-scale LHC measurements are phenomenologically relevant~\cite{Niemi:2018asa,Ramsey-Musolf:2019lsf,Gould:2019qek,Chala:2018opy}.

Along these lines, two Higgs doublet models (2HDMs) remain attractive
theories; they have seen continued scrutiny in the
literature~\cite{Dorsch:2014qja,Dorsch:2016tab,Goncalves:2021egx,Su:2020pjw,Dorsch:2016nrg,Wang:2021ayg,Basler:2016obg,Dorsch:2017nza,Atkinson:2021eox,Basler:2021kgq}. On
the one hand, currently available experimental results are not
sensitive enough to move exotic scalar bosons beyond the kinematic
reach of the LHC~\cite{Atkinson:2022pcn}. On the other hand, 
electroweak precision constraints 
are 
avoided similar 
as in the Standard Model.\footnote{It should be noted that the
  observed tension in the Kaon sector~\cite{Atkinson:2021eox} and the
  recently experimentally hardened $(g-2)_\mu$
  observation~\cite{Athron:2021iuf,Atkinson:2021eox} remain difficult
  to explain in the minimal implementations of the 2HDM.} One
important shortfall of 2HDMs, in particular in its type II
manifestation that provides a tangible link to supersymmetric UV
completions of the SM, is the generic difficulty of obtaining a strong first-order electroweak phase transition (EWPT) for existing parameter
constraints~\cite{Basler:2016obg,Basler:2018dac,Basler:2019iuu,Atkinson:2021eox}. As
2HDM dynamics alone do not seem to be quite enough to furnish a strong first-order
EWPT, it is the purpose of this paper to clarify the extra dynamics
that are required for the 2HDM to provide a sufficiently large EWPT
for electroweak baroygenesis. Concretely, we approach this by means of
effective field theory (see
also~\cite{Balazs:2016yvi,deVries:2017ncy,Croon:2020cgk}) and focus in
this work on extensions of the scalar potential of the
softly broken $\mathbb{Z}_2$-symmetric and CP-conserving
2HDM as a well-motivated sector to facilitate a strong first-order
EWPT~\cite{Grojean:2004xa,Ham:2004zs,Bodeker:2004ws,Zhang:1992fs,Profumo:2007wc,Morrissey:2012db}. We
will focus on the 2HDM type II in this work, but as we will focus
mostly on the implications for multi-Higgs production and
phenomenological prospects for multi-top final states, our findings
generalise to the 2HDM type I straightforwardly.

We organise this work as follows: In Sec.~\ref{sec:2hdmeft} we review the basics of the 2HDM alongside the effective field theory (EFT) modifications we consider in this work. Section~\ref{sec:ewpt} provides a short overview of our computational methods. Section~\ref{sec:results} is devoted to our results: we provide scans of operators to achieve a strong first-order EWPT and clarify the correlated phenomenological implications relevant for the LHC in multi-Higgs and multi-top final states. We summarise and conclude in Sec.~\ref{sec:conc}.

\section{2HDMs and Dimension-6 Higgs Potential Extensions}
\label{sec:2hdmeft}
The tree-level dimension-4 potential of the 2HDM is given
by~\cite{Gunion:1989we,Gunion:2002zf} 
\begin{multline}
\label{eq:2HDMsc}
 V_{\text{tree}}(\Phi_1, \Phi_2) = {m}^2_{11}(\Phi_1^{\dagger}\Phi_1) + {m}^2_{22}(\Phi_2^{\dagger}\Phi_2) - {m}^2_{12}(\Phi_1^{\dagger}\Phi_2 + \Phi_2^{\dagger}\Phi_1) + {{\lambda}_1}(\Phi_1^{\dagger}\Phi_1)^2 + {{\lambda}_2}(\Phi_2^{\dagger}\Phi_2)^2  \\
	 +  {\lambda}_3(\Phi_1^{\dagger}\Phi_1)(\Phi_2^{\dagger}\Phi_2) + {\lambda}_4(\Phi_1^{\dagger}\Phi_2)(\Phi_2^{\dagger}\Phi_1) +
	\frac{1}{2}{\lambda}_5[(\Phi_1^{\dagger}\Phi_2)^2 + (\Phi_2^{\dagger}\Phi_1)^2]  \\
	 +  \left({\lambda}_6(\Phi_1^{\dagger}\Phi_1) +{\lambda}_7(\Phi_2^{\dagger}\Phi_2)\right)\left(\Phi_1^{\dagger}\Phi_2 + \Phi_2^{\dagger}\Phi_1\right)
\end{multline}
where $\Phi_{1,2}$ are $SU(2)_L$ doublets with hypercharge
$Y=1$. The absence of tree-level flavour-changing neutral
interactions can be guaranteed by imposing a $\mathbb{Z}_2$ symmetry~\cite{Glashow:1976nt},
which is softly broken by the term proportional $m_{12}^2$. In the
following, we will assume only such soft breaking and choose the
couplings $\lambda_{6,7}=0$, which induces a hard breaking of
$\mathbb{Z}_2$, to be zero. We furthermore take the values of the
remaining coupling and mass parameters, $\lambda_i$ ($i=1,...,5$) and
$m_{ab}^2$ ($a,b=1,2$), to be real.

\begin{table}[t!]
	\centering
	\renewcommand{\arraystretch}{1.8}
    \begin{tabular}{|c|c||c|c|}
			\hline \hline
			$O_6^{111111}$&
			$(\Phi_1^{\dagger}\Phi_1)^3$ & 
			$O_6^{222222}$&
			$(\Phi_2^{\dagger}\Phi_2)^3$ \\
			
			$O_6^{111122}$&
			$(\Phi_1^{\dagger}\Phi_1)^2(\Phi_2^{\dagger}\Phi_2)$&
			$O_6^{112222}$&
			$(\Phi_1^{\dagger}\Phi_1)(\Phi_2^{\dagger}\Phi_2)^2$ \\
			
			$O_6^{122111}$&
			$(\Phi_1^{\dagger}\Phi_2)(\Phi_2^{\dagger}\Phi_1)(\Phi_1^{\dagger}\Phi_1)$&
			$O_6^{122122}$&
			$(\Phi_1^{\dagger}\Phi_2)(\Phi_2^{\dagger}\Phi_1)(\Phi_2^{\dagger}\Phi_2)$\\
			
			$O_6^{121211}$&
			$(\Phi_1^{\dagger}\Phi_2)^2(\Phi_1^{\dagger}\Phi_1)$ + h.c. &
			$O_6^{121222}$&
			$(\Phi_1^{\dagger}\Phi_2)^2(\Phi_2^{\dagger}\Phi_2)$ + h.c. \\
			\hline \hline
	\end{tabular}
	\caption{Dimension-6 operators of class $\Phi^6$  involving  $\Phi_1$ and $\Phi_2$.}\label{tab:phi6op}
\end{table}    

Including higher-dimensional EFT contributions to the Higgs potential leads to the operators of Tab.~\ref{tab:phi6op}. The $\Phi^6$ operators\footnote{ The dimension-6 operators are classified following the Warsaw basis convention \cite{Grzadkowski:2010es}. } that we focus on in this first investigation give rise to a dimension-6 extension of the 2HDM potential 
\cite{Anisha:2019nzx,Crivellin:2016ihg,Banerjee:2020bym,Karmakar:2017yek}
\begin{equation}
	{\cal{L}}_{\text{EFT}} = {\cal{L}}_{\text{2HDM}} + \sum_{i} 
	{C^{i}_{6}\over \Lambda^2 } O^{i}_{6}
	\quad \Rightarrow \quad
	V_{\text{dim-6}} = -  \sum_{i}
	{C^{i}_{6}\over \Lambda^2} O^{i}_{6} \;. \label{eq:dim6pot}
\end{equation}
Here, $ O^{i}_{6}$ are the dimension-6 operators given in Tab.~\ref{tab:phi6op} and $C^{i}_{6}$ are the corresponding Wilson Coefficients (WCs).
As in the unperturbed 2HDM we can solve the tadpole equations to relate $m_{11}^2$ and $m_{22}^2$ to the remaining potential parameters in the vacuum. Diagonalisation of the charged and CP-odd Higgs mass mixing matrices is described by the characteristic angle
\begin{equation}
\tan \beta =  {v_2\over v_1}\,,
\end{equation}
which means that even in the presence of our $\Phi^6$ interactions
$\tan\beta$ is directly linked to the ratio of vacuum expectation
values, while $(246~\text{GeV})^2\approx v^2 = v_1^2+v_2^2$ is fixed by the $W$
boson mass (or equivalently the Fermi constant). Explicit expressions
for the masses of the neutral and charged Higgs bosons can be obtained
similar as in~\cite{Anisha:2019nzx}, however, it is convenient for us
to choose masses and mixing angles as input parameters. Taking
inspiration from the on-shell renormalisation scheme introduced in~\cite{Basler:2016obg}, we perform
shifts of the renormalisable part of the Lagrangian, $\lambda_i
\to \lambda_i + \delta \lambda_i$ and $m_{12}^2 \to m_{12}^2 + \delta
m_{12}^2$, with
\begin{equation}
\label{eq:lambdashift}
\begin{split}
\delta \lambda_1^{\text{d6}} =& {1\over  4 \Lambda^2 v_1^2 } \big[6 C_6^{111111}v_1^4  + (2 C_6^{121211} + C_6^{122111}) v_1^2v_2^2 \\& - 
  \{2 (C_6^{112222} + C_6^{121222}) + C_6^{122122})v_2^4\}\big]\,,\\
 \delta \lambda_2^{\text{d6}} =& -{1\over 4\Lambda^2 v_2^2} \big[ \{2(C_6^{111122} + C_6^{121211}) + C_6^{122111}\}v_1^4 \\
& - (2C_6^{121222} + C_6^{122122})v_1^2 v_2^2 - 
   6C_6^{222222} v_2^4 \big]\,,\\
\delta \lambda_4^{\text{d6}} =& {v_1^2\over \Lambda^2} (C_6^{111122} + C_6^{121211} + C_6^{122111}) \\ & + {v_2^2\over \Lambda^2}(C_6^{112222} + C_6^{121222} + C_6^{122122})\,,\\
\delta \lambda_5^{\text{d6}} =&{1\over 2 \Lambda^2}\big[(2C_6^{111122} + 4C_6^{121211} + C_6^{122111})v_1^2 \\& + (2C_6^{112222} + 4C_6^{121222} + C_6^{122122}) v_2^2\big]\,,\\
\delta m_{12}^{2\,{\text{d6}}} =& {v_1 v_2\over 2\Lambda^2} \big[\{2(C_6^{111122} + C_6^{121211}) + C_6^{122111}\} v_1^2 \\ & + \{2(C_6^{112222} + C_6^{121222}) + C_6^{122122}\} v_2^2\big]\,.
\end{split}
\end{equation} 
They directly yield mass eigenvalues and mixing angles as in the $d=4$ 2HDM.\footnote{A similar consistent choice could be obtained by a different subset of the potential parameters in the dimension-4 Lagrangian.} 
In the following, we will refer to $h$ and $H$ as the
  lighter and the heavier CP-even Higgs boson, respectively. $A$ denotes the CP-odd scalar and $H^+$ is the charged scalar degree of freedom.
This means that for any choice of Wilson coefficients we obtain the
same mass spectrum and neutral mixing angle in the vacuum as for
Eq.~\eqref{eq:2HDMsc}. Any direct coupling of the Higgs bosons to SM
matter is therefore insensitive to the Wilson coefficients and the
choice of Eq.~\eqref{eq:lambdashift} shifts correlations into Higgs
self-couplings and multi-Higgs final states, given single Higgs
measurements as the transparent input that is typically provided by
experimental collaborations and checked for in parameter scans with
BSM tools such as
{\tt{ScannerS}}~\cite{Coimbra:2013qq,ScannerS,Muhlleitner:2020wwk},
{\tt{HiggsBounds}}~\cite{Bechtle:2008jh,Bechtle:2011sb,Bechtle:2013wla,Bechtle:2020pkv},
or
{\tt{HiggsSignals}}~\cite{Bechtle:2013xfa,Bechtle:2020uwn}.\footnote{The
  dimension-4 parameters can then be obtained by inverting $\delta
  \lambda_i,\delta m_{12}^2$ in the $\Lambda^{-1}$ expansion.}

Consequently, the usual classification of the 2HDM according to the
$\mathbb{Z}_2$ assignments applies to this work as well. The Higgs
boson couplings to fermions $f$ in the mass basis are given by
\begin{multline}
\label{eq:yuk}
 {\cal L}_{\text{Yuk}} = 
 - \sum \limits_{f= u,d,\ell} \frac{m_f}{v} \left(\xi_h^f \, \bar{f} f h + \xi_H^ f \, \bar{f} f H - i \xi_A^f  \, \bar{f} \gamma_5 f A \right) \\
 -  \left[ \frac{\sqrt{2} V_{ud}}{v} \bar{u} \left(m_d \, \xi_A^d {\text{P}}_{\text{R}} - m_u \, \xi_A^u {\text{P}}_{\text{L}} \right) d H^+ + \frac{\sqrt{2}}{v} m_\ell \, \xi_A^l  (\bar{\nu} {\text{P}}_{\text{R}} \ell)  H^+ + {\text h.c.}\right]\,,
\end{multline}
where ${\text{P}}_{\text{L,R}}$ are the left and right chirality projectors and the coupling modifiers $\xi$ are given in 
Tab.~\ref{tab:coupmod}.

\begin{table}[!t]
    \centering
    \begin{tabular}{|c||c|c|c|c|}
        \hline
        \hline
        \rm Model & I & II \\
        \hline
        $\xi_h^u$ & $\cos\alpha/\sin\beta$ & $\cos\alpha/\sin\beta$  \\
        \hline
        $\xi_h^d$ & $\cos\alpha/\sin\beta$ & $-\sin\alpha/\cos\beta$  \\
        \hline
        $\xi_H^u$ & $\sin\alpha/\sin\beta$ & $\sin\alpha/\sin\beta$ \\
        \hline
        $\xi_H^d$ & $\sin\alpha/\sin\beta$ & $\cos\alpha/\cos\beta$  \\
        \hline
        $\xi_A^u$ & $\cot\beta$ & $\phantom{-}\cot\beta$  \\
        \hline
        $\xi_A^d$ & $\cot\beta$ & $-\tan\beta$  \\
        \hline
        \hline
    \end{tabular}
    \caption{Coupling modifiers $\xi$ for 2HDM type I and II and up- and down-type quarks relevant for this study.}
    \label{tab:coupmod}
\end{table}

\section{Finite Temperature and Phase Transitions}
\label{sec:ewpt}
In order to calculate the strength of the EWPT in our model, we determine the global minimum of the one-loop corrected effective potential at finite temperature. The derivation of the effective potential is reviewed in Sec.~\ref{subsec:calcmeth}, our scan is described in Sec.~\ref{subsec:scans}.
\subsection{Review of Calculational Methods}\label{subsec:calcmeth}
The effective potential describes the exact vacuum state~\cite{Coleman:1973jx} of a theory including finite temperature effects~\cite{Dolan:1973qd,Carrington:1991hz,Quiros:1999jp}. It can be derived through a perturbative expansion of the generating functional of one-particle irreducible Green's functions. Consequently, the one-loop contribution includes the inverse propagator, independent of the structure of the underlying Lagrangian.
In terms of the static field configurations described
  by $\vec{\omega}$,
  and the temperature $T$, the one-loop contribution to the effective potential at finite
temperature has the general form 
\begin{align}
    V_\text{eff}^{(1)} (\vec{\omega},T) =\sum_{X=S,G,F} (-1)^{2s_X}
  (1+2s_X) I^X \,,
\end{align}
\allowdisplaybreaks
with scalar $(S)$, gauge-boson $(G)$ and fermion $(F)$ contributions that have the following form
\begin{subequations}
    \label{eq:1LoopInts}
    \begin{align}
        I^S &= \frac{T}{2}\sum_n^\text{Bos}\int\frac{d^3 k}{(2\pi)^3} \sum_i \left[\log \det\left(-\mathcal{D}^{-1}_{S,\, i}\right)\right]\,,\\
        I^{G} &= \frac{T}{2}\sum_n^\text{Bos}\int\frac{d^3 k}{(2\pi)^3} \sum_i \left[\log \det\left(-\mathcal{D}^{-1}_{GB,\, i}\right)\right]\,,\\
        I^F &= -T \sum_n^\text{Ferm}\int\frac{d^3 k}{(2\pi)^3} \sum_i \left[\log \det\left(-\mathcal{D}^{-1}_{F,\, i}\right)\right]\,,
    \end{align}
\end{subequations}
with inverse propagators $\mathcal{D}^{-1}$ calculated in finite
temperature field theory. Within the imaginary time formalism, the
propagators receive temperature-dependent corrections, {\it{e.g.}} the
inverse scalar propagator in momentum space has the form
$\mathcal{D}_S^{-1}=\omega_n^2+\omega_k^2$ in terms of
the discrete Matsubara modes
\begin{align}
    \omega_n^2 &= (2n\pi T)^2\,,\quad n\in\mathbb{N}_0
\end{align}
and
\begin{align}
    \omega_k^2 &= \bm{k}^2 + m^2\label{eq:omegak}\,,
\end{align}
where in the case of Eq.~\eqref{eq:omegak}, the mass term receives corrections from the EFT operators. 
At $T=0$ this gives rise to the vacuum expectation value $v\simeq
246~\text{GeV}$ as described above. 
 
The integrals in Eq.~\eqref{eq:1LoopInts} split into a UV-divergent
temperature-independent part and a UV-finite, but IR-divergent
temperature-dependent part. In the $\overline{\text{MS}}$-scheme, the
one-loop contributions read 
\begin{align}
    I^X_{\overline{MS}} = \underbrace{\frac{m_X^4}{64 \pi^2}\left[\log\left(\frac{m_X^2}{\mu^2}\right)-k_X\right]}_{\equiv V_\text{CW}(\vec{\omega})} + \underbrace{\frac{T^4}{2\pi^2} J_\pm\left(\frac{m_X^2}{T^2}\right)}_{\equiv V_T(\vec{\omega},T)}
\end{align}
with $X=\{(S), (G), (F)\}$ and the renormalisation constant $k_X$
\begin{align}
    k_X=\begin{cases}
        \frac{5}{6},\quad\text{for gauge bosons}\\
        \frac{3}{2},\quad\text{otherwise}
    \end{cases}
\end{align}
and the \textit{thermal} fermionic $(+)$ and bosonic $(-)$ function $J_\pm$~\cite{Dolan:1973qd,Quiros:1994dr,Quiros:1999jp}
\begin{align}
    J_{\pm}(x^2) = \int_0^\infty dk k^2 \log\left(1 \pm e^{-\sqrt{k^2+x^2}}\right)\,.
\end{align}
The bosonic Matsubara zero modes, $n = 0$, lead to
IR divergences that are 
cancelled by resumming the thermal masses $\Pi$
\cite{Carrington:1991hz,Parwani:1991gq,Arnold:1992fb,Kapusta:2006pm,Arnold:1992rz,Quiros:1994dr}. The
scalar thermal masses are calculated as thermal scalar self-energy
corrections in the soft-momentum limit,
\begin{align}
  \nonumber
  \Pi_{ij}^{(1)}(\bm{p}\rightarrow 0,\omega_n \rightarrow 0) &\equiv \Pi_{ij}^{(1)}(0)\\ 
  &= \sum_k \kappa_{ij}^k T\sum_n\int \frac{d^3 p}{(2\pi)^3} \mathcal{D}_{kk}(\omega_n,\omega_p) \\
  &\quad+ \sum_{k,l} \kappa_{ij}^{kl} T^2\sum_{n,m}\int \frac{d^3 p_1}{(2\pi)^3} \mathcal{D}_{kk}(\omega_n,\omega_{p_1})\int \frac{d^3 p_2}{(2\pi)^3} \mathcal{D}_{ll}(\omega_m, \omega_{p_2})\nonumber\,.
\end{align}
Here, quartic scalar couplings are labelled with $\kappa_{ij}^k$, while couplings between six scalars are encoded in $\kappa_{ij}^{kl}$. 
Note, that the latter is already a two-loop correction to the scalar self-energy.
The finite temperature field theory integral is evaluated in the
high-temperature limit $m/T\ll 1$ as
\begin{multline}
    T\sum_n\int \frac{d^3 p}{(2\pi)^3} \mathcal{D}_{ij}(\omega_n, \omega_p) = T\sum_n\int \frac{d^3 p}{(2\pi)^3} \frac{1}{\omega_n^2 + \omega_p^2} \\ = T\sum_n\int \frac{d^3 p}{(2\pi)^3} \frac{1}{\omega_n^2 + \bm{p}^2 + m^2} \simeq T\sum_n\int \frac{d^3 p}{(2\pi)^3} \frac{1}{\omega_n^2 + \bm{p}^2}
    = \dots = \frac{T^2}{12}\left[1+\mathcal{O}\left(\frac{m}{T}\right)\right]\,.\label{eq:integral}
\end{multline}
The dimension-6 operators generate 2-loop contributions $\sim T^4$ (see
{\it{e.g.}}~\cite{Bodeker:2004ws}) which we include throughout our
calculation. These derive straightforwardly from the six-point
interaction vertices of Tab.~\ref{tab:phi6op}. Applying the
Arnold-Espinosa method \cite{Arnold:1992rz}\footnote{For 
  further remarks on this approach and how it compares to the Parwani
  approach, {\it{cf.}}~\cite{Arnold:1992rz,Parwani:1991gq}. Further
  discussions and comparisons are given in
  \cite{Cline:1996mga,Cline:2011mm}.}, the thermal potential 
$V_T (\vec{\omega},T)$ is replaced as
\begin{eqnarray}
V_T (\vec{\omega},T) \to V_T (\vec{\omega},T) + V_{\text{daisy}}
(\vec{\omega},T) \;,
\end{eqnarray}
where
\begin{eqnarray}
V_{\text{daisy}} (\vec{\omega},T) = - \frac{T}{12\pi} \left[
  \sum_{i=1}^{n_{\text{Higgs}}} \left( (\overline{m}_i^2)^{3/2} -
    (m_i^2)^{3/2} \right) + \sum_{a}^{n_{\text{gauge}}} \left(
    (\overline{m}_a^2)^{3/2} - (m_a^2)^{3/2} \right) \right] \,,
\end{eqnarray}
with $n_{\text{Higgs}}$ denoting the number of real Higgs fields and
$n_{\text{gauge}}$ the number of gauge bosons in the adjoint
representation of the gauge group. The $\overline{m}$ denote the thermal
masses that include the thermal corrections $\Pi^{(1)}$. 

As stated above, we introduce shifts to the parameters of the Higgs
potential to absorb the effect of the dimension-6 operators such that
the masses and mixing angles that we use as input parameters remain
unchanged. We also require the one-loop corrections to leave the
masses and mixing angles at their tree-level values. For this we
introduce additional finite counterterms summarised in the counterterm
potential $V^{\text{CT}}$ and apply the renormalisation conditions 
\begin{eqnarray}
0 &=& \partial_{\phi_i}
(V^{\text{CW}}+V^{\text{CT}}|_{\vec{\bar{\omega}}=\vec{\bar{\omega}}_{\text{tree}}})
\\
0 &=& \partial_{\phi_i} \partial_{\phi_j}
(V^{\text{CW}}+V^{\text{CT}}|_{\vec{\bar{\omega}}=\vec{\bar{\omega}}_{\text{tree}}}) \;,
\end{eqnarray}
where $\phi_i$ denote the scalar Higgs doublet fields developing a
non-zero VEV $\bar{\omega}_i$, with $\bar{\omega}_{\text{tree},i}$
being the corresponding tree-level VEV.
Note, that for
the construction of our counterterm potential, we choose
the free parameters arising in the derivation of the potential,
{\it{cf.}}~\cite{Basler:2016obg}, such that the dimension-6 terms 
do not introduce new coupling structures. The dimension-6 operator
contributions enter through the partial derivatives of
$V_{\text{CW}}$ which, however, are modified w.r.t.~the 2HDM through the dimension-6
contributions to the mass terms.
Note, that the dimension-6
operators do not introduce additional new counterterm structures in
the above two conditions. 
For further details on the renormalisation
procedure, we refer to~\cite{Basler:2018cwe,Basler:2020nrq}.

With the dimension-6 extended tree-level potential
  $V_{\text{tree,dim-6}}$, 
\begin{eqnarray}
V_{\text{tree,dim-6}} \equiv V_{\text{tree}} + V_{\text{dim-6}} \,
\end{eqnarray}
in terms of the tree-level potential
  $V_{\text{tree}}$ of Eq.~(\ref{eq:2HDMsc}) and the dimension-6 potential
  $V_{\text{dim-6}}$ of Eq.~(\ref{eq:dim6pot}), we hence have for the
  loop-corrected effective potential at finite temperature as function
  of the classical field configuration $\vec{\omega}$,
\begin{eqnarray}
V(\vec{\omega},T) = V_{\text{tree,dim-6}} (\vec{\omega}) +
V_{\text{CW}} (\vec{\omega}) + V_{\text{CT}} (\vec{\omega}) + V_T
(\vec{\omega},T) \;,
\end{eqnarray}
which we have implemented in the C++ code {\tt
  BSMPT}~\cite{Basler:2018cwe,Basler:2020nrq}.

For an electroweak phase transition to be of strong first order, the
ratio of the critical VEV $v_c$ at the critical temperature $T_c$ has
to fulfil the (conventionally chosen) criterion $\xi_c \equiv v_c/T_c >1$ to avoid too large 
baryon washout. 
The critical temperature is defined as
the temperature where there exist two degenerate global minima, one at
$v=0$ and the other at the critical VEV $v_c\neq 0$. The values of $T_c$,
$v_c$ and hence $\xi_c$ are obtained from {\tt BSMPT} which computes
the vacuum expectation value $v(T)$ at a given temperature $T$ through the
minimisation of the effective potential $V(\vec{\omega},T)$. The
value $v$ is obtained as
\begin{eqnarray}
v(T) = \left( \sum_{k=1}^{n_H} \bar{\omega}_k^2 \right)^{\frac{1}{2}} \;,
\end{eqnarray}
where $n_H$ means that the sum is performed over all field directions
in which we allow for the development of a non-zero electroweak VEV,
which are given by the fields that couple to the electroweak gauge
bosons. 
The $\bar{\omega}_k$ denote the field configurations that
minimise the loop-corrected effective potential $V(\vec{\omega},T)$.

\subsection{Scanning Methodology}
\label{subsec:scans}
For our numerical analysis, we use parameter points that are compatible
with all relevant theoretical and experimental constraints. We resort
here to a parameter sample that has been generated recently for the
2HDM \cite{Abouabid:2021yvw}. The scan was performed with the help of
the program {\tt ScannerS} 
\cite{Coimbra:2013qq,ScannerS,Muhlleitner:2020wwk}. {\tt ScannerS}
chooses as scan parameters the masses of the 2HDM Higgs
bosons, $\tan\beta$, the soft breaking mass term $m_{12}^2$ and the
coupling $c_{HVV}$ of the heavier Higgs boson to massive gauge bosons
$V\equiv W^\pm, Z$ (instead of the mixing angle $\alpha$ that
diagonalizes the neutral CP-even mass matrix). We
  performed an additional scan to select parameter points with
  significant branching ratios $\mbox{BR}(H\to hh)$ of $H$ into a pair
  of lighter Higgs bosons $h$. The total scan
ranges of the two merged parameter samples are listed in
Tab.~\ref{tab:r2hdmranges} for the scenario where the 
lighter of the two neutral CP-even Higgs bosons, $h$, takes the role of
the SM-like Higgs, denoted as $H_{\text{SM}}$ in the
following.\footnote{While also scenarios with $H\equiv H_{\text{SM}}$
  still comply with all applied constraints, the plots shown in the
  following have $h \equiv H_{\text{SM}}$.} 
We restrict ourselves to the type
I and II models. {\tt ScannerS} checks for the theoretical
constraints, requiring that the potential is bounded from below,
that perturbative unitarity holds and that the electroweak vacuum is
the global minimum.  For the latter it uses the discriminant from
\cite{Barroso:2013awa}. 

\renewcommand{\arraystretch}{1.2}
\begin{table}[b!]
\begin{center}
\begin{tabular}{|c|c|c|c|c|c|c|}
\hline
$m_{h}$ [GeV] & $m_{H}$ [GeV] & $m_A$ [GeV] & $m_{H^\pm}$ [GeV] & $\tan\beta$ &
$c_{H VV}$ & $m_{12}^2$ [GeV$^2$] \\ \hline \hline
\multicolumn{7}{|c|}{2HDM I/II (light)} \\ \hline
125.09 & 130...3000 & 30...3000 & 85/800...3000 & 0.8...30 & -0.3...1.0 &
 $10^{-5}$...$10^7$ \\ \hline
\end{tabular}
\end{center}
\vspace*{-0.3cm}
\caption{Scan ranges of the 2HDM input parameters, where light refers
to the set-up where the lighter of the two CP-even neutral Higgs
bosons is the SM-like Higgs $H_{\text{SM}}$, {\it i.e.}~$h \equiv
H_{\text{SM}}$.  
\label{tab:r2hdmranges} }
\end{table}
\renewcommand{\arraystretch}{1} 

On the experimental side, we impose compatibility with the
electroweak precision data and demand the computed $S$,
$T$ and $U$ values to be within $2\sigma$ of the SM fit
\cite{Baak:2014ora}, taking into account the full 
correlation among the three parameters.
One of the neutral CP-even Higgs bosons, in our study
chosen to be $h$, is required to have a mass of
\cite{ATLAS:2015yey} 
\begin{eqnarray}
m_{H_{\text{SM}}} = 125.09 \, \mbox{GeV} \,,
\end{eqnarray}
and to behave SM-like. {\tt Scanners} checks for compatibility with the Higgs signal data 
through the link to {\tt HiggsSignals} version 2.6.1
\cite{Bechtle:2013xfa}. Scenarios with interfering Higgs signals are
excluded by forcing the non-SM-like Higgs bosons to lie outside an
interval of 5~GeV around 125~GeV in order to avoid interference
effects that require a dedicated thorough study beyond the focus of this work.
We require 95\% C.L. exclusion limits on non-observed scalar states by using {\tt
HiggsBounds} version 5.9.0 \cite{Bechtle:2008jh,Bechtle:2011sb,Bechtle:2013wla,Bechtle:2020pkv}.
The sample is also checked with respect to the recent ATLAS
  analyses in the $ZZ$ \cite{ATLAS:2020tlo} and $\gamma\gamma$
  \cite{ATLAS:2020tws} final 
  states that were not yet included in {\tt HiggsBounds}.
Flavour constraints are taken into account by testing for
compatibility 
with $\mathcal{R}_b$ \cite{Haber:1999zh,Deschamps:2009rh} and
$B\rightarrow X_s \gamma $
\cite{Deschamps:2009rh,Mahmoudi:2009zx,Hermann:2012fc,Misiak:2015xwa,Misiak:2017bgg, Misiak:2020vlo} in the
$m_{H^{\pm}}-\tan\beta$ plane. We imposed in the 2HDM type II the
latest bound on the charged Higgs mass given in \cite{Misiak:2020vlo},
$m_{H^\pm} \ge 800$~GeV, for essentially all values of $\tan\beta$. In
the type I models, this bound is much weaker and is strongly correlated
with $\tan\beta$. 

The inclusion of the Wilson coefficients in the 2HDM potential
modifies the 2HDM mass values and mixing angles. By applying the
shifts given in Eqs.~(\ref{eq:lambdashift}) we ensure, however,
that they remain unchanged also after
inclusion of the dimension-6 contributions. This allows
us to use the {\tt ScannerS} sample of 
parameter points and work with parameter sets that are compatible with
all relevant theoretical and experimental constraints. In practice, we
apply within {\tt BSMPT} on the parameters $\lambda_{1,2,4,5}$ and
$m_{12}^2$ of the respective {\tt ScannerS} sample point the shifts of
Eqs.~(\ref{eq:lambdashift}). The values of $m_{11}^2$ and $m_{22}^2$ are obtained from
the minimisation conditions taking into account the dimension-6
contributions to the Higgs potential. With these parameters and
$\lambda_3$ for the given parameter point under investigation, {\tt
  BSMPT} then computes the EWPT for the 2HDM potential including the
dimension-6 operators. 

\section{Phenomenological Aspects of Effective 2HDM Phase Transitions}
\label{sec:results}
The general requirements for EFT methods to provide appropriate
approximations for momentum-independent Wilson coefficients are
twofold. Firstly, the heavy degrees of freedom that are integrated out
need to be sufficiently heavy compared to the characteristic energy
scale that is probed at the respective laboratory. And secondly,
perturbativity, which constitutes the overarching concept of the
field-theoretic aspects investigated in this work imposes the
additional requirement of dimension-6 terms to be a small correction
in relation to the renormalisable dimension-4 result. This provides
confidence in neglecting higher order terms in the $\Lambda^{-1}$
expansion. 

The modifications of the 2HDM introduced in Sec.~\ref{sec:2hdmeft}
provide a rich landscape for phenomenological deviations. This is
particularly interesting for the 2HDM type II that we will mostly
focus on in this section. However, as we are mainly considering
interactions of the extra Higgs bosons with the top sector, our
findings generalise to the 2HDM type I, where a strong first-order
electroweak phase transition (SFOEWPT) already at
  dimension-4 level can be found more easily than in type II.
\begin{figure*}[!t]
\centering
\includegraphics[width=0.43\textwidth]{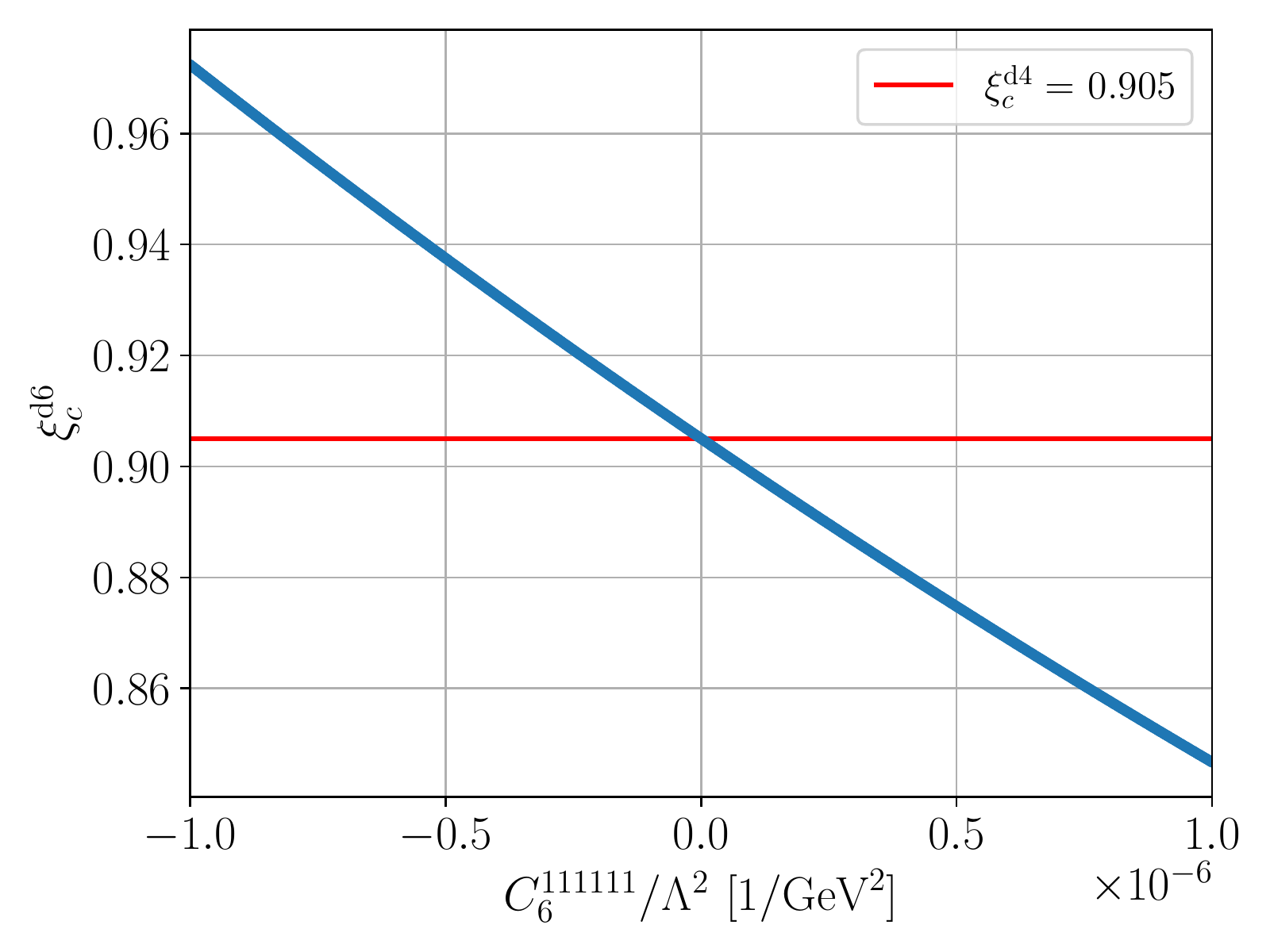}\hspace{0.6cm}
\includegraphics[width=0.43\textwidth]{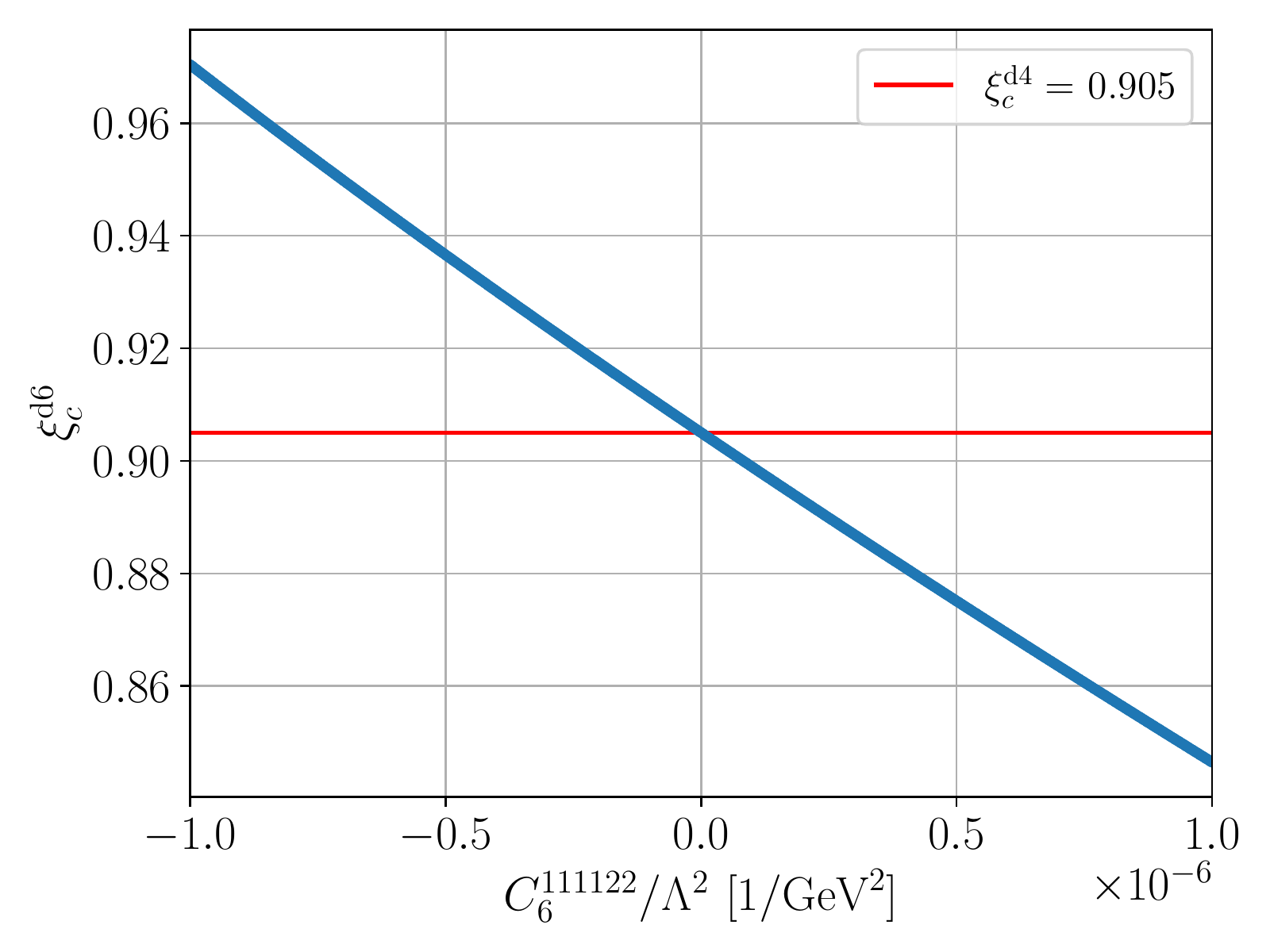}\\
\includegraphics[width=0.43\textwidth]{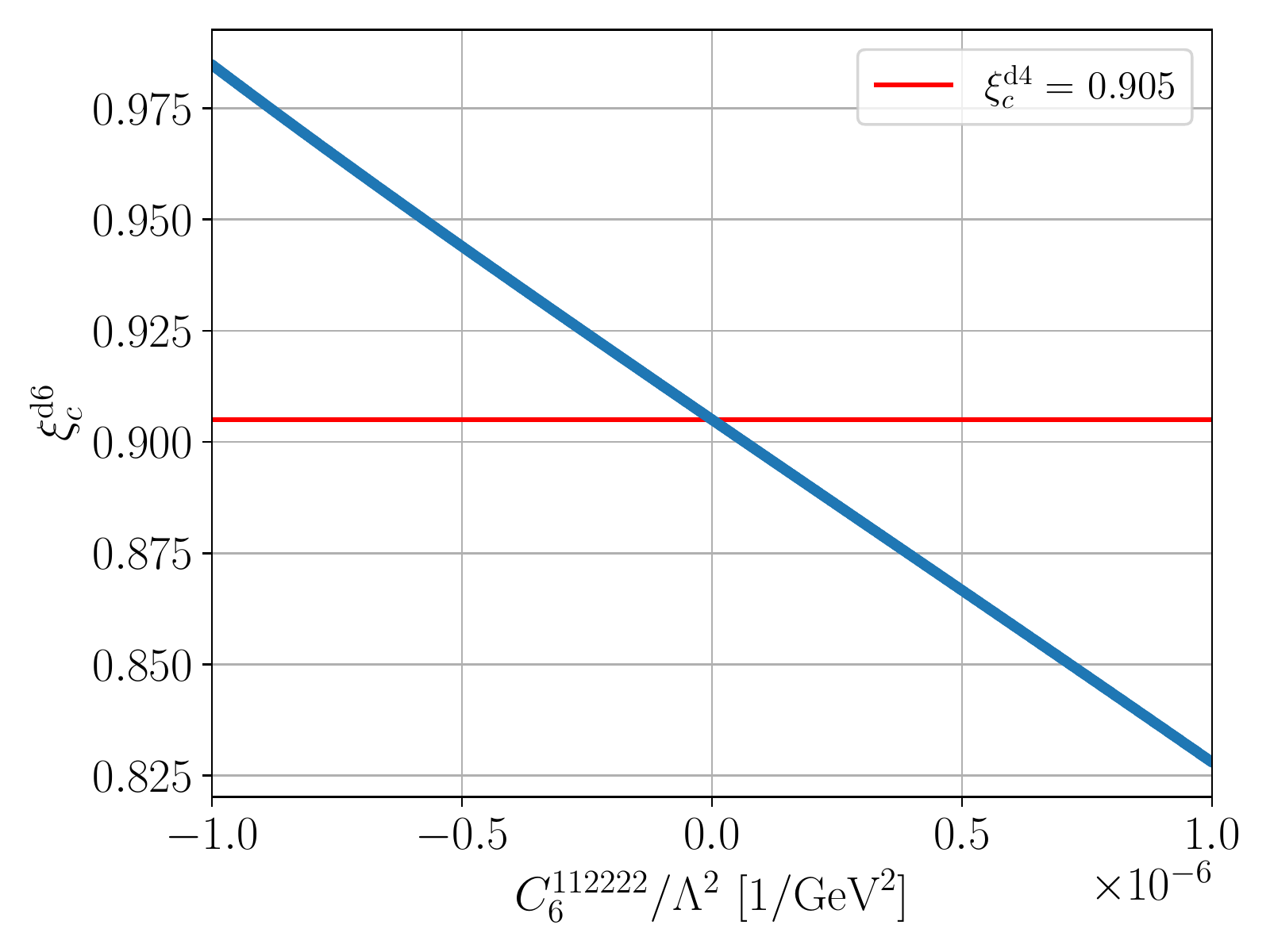}\hspace{0.6cm}
\includegraphics[width=0.43\textwidth]{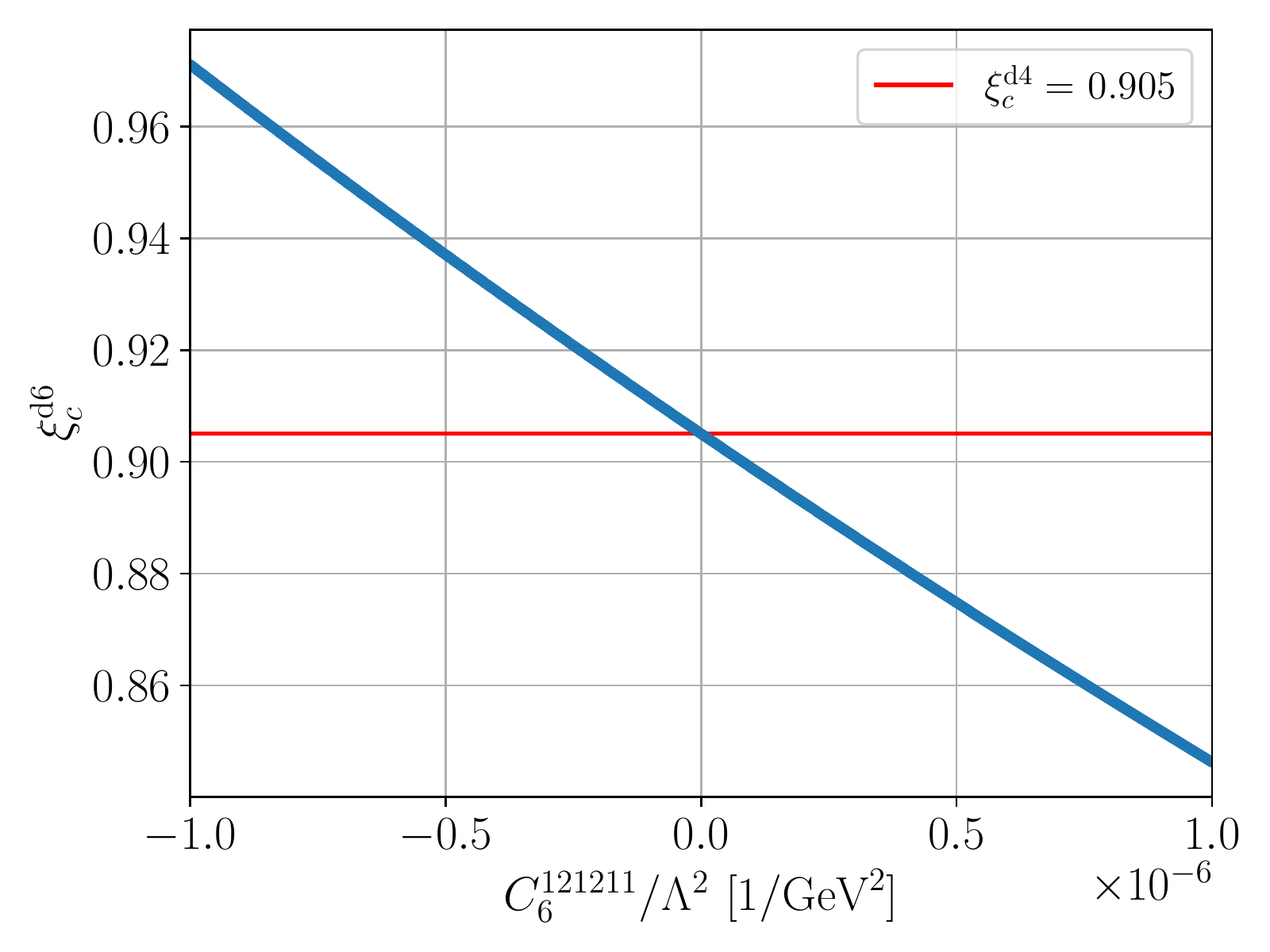}\\
\includegraphics[width=0.43\textwidth]{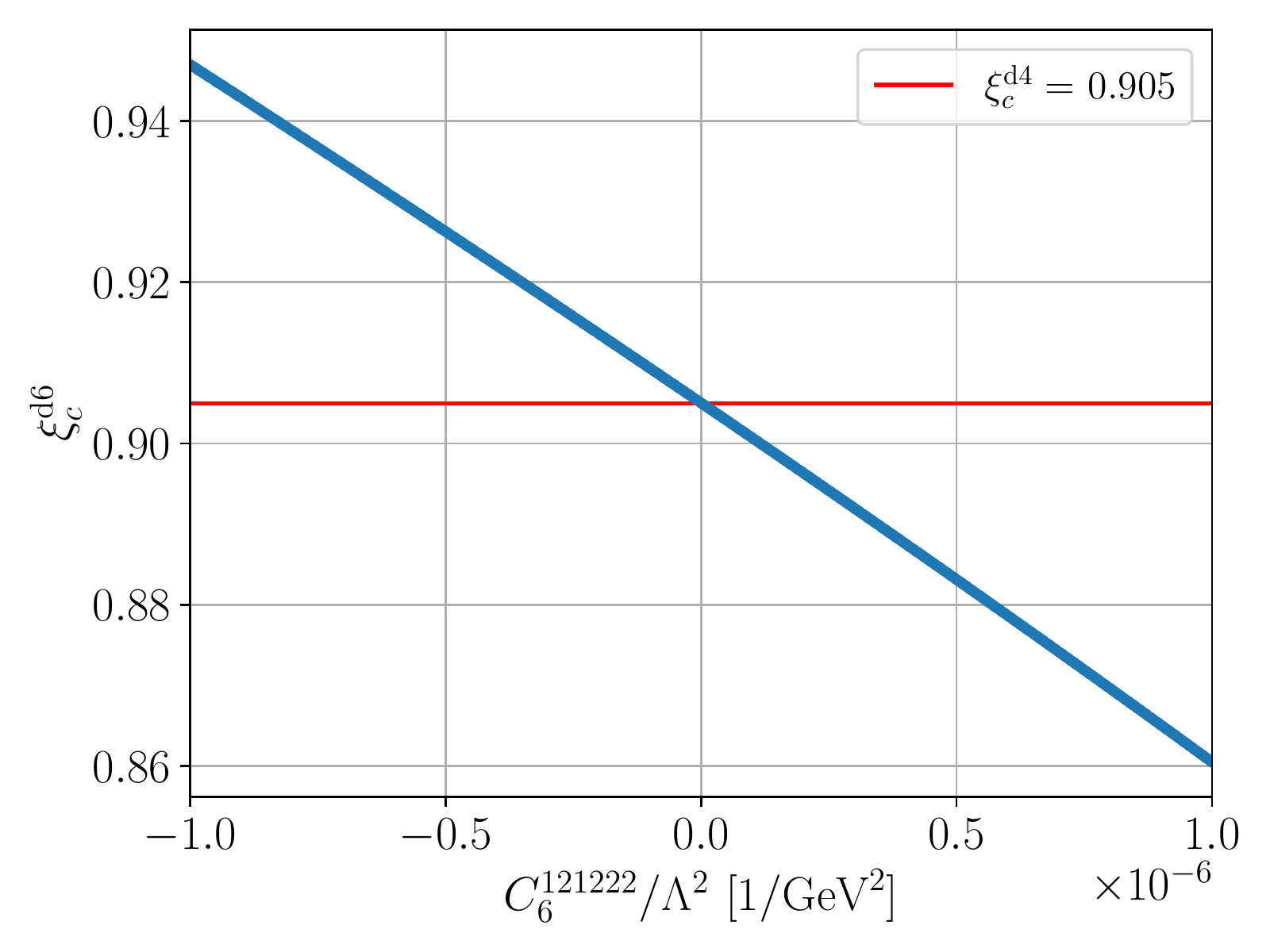}\hspace{0.6cm}
\includegraphics[width=0.43\textwidth]{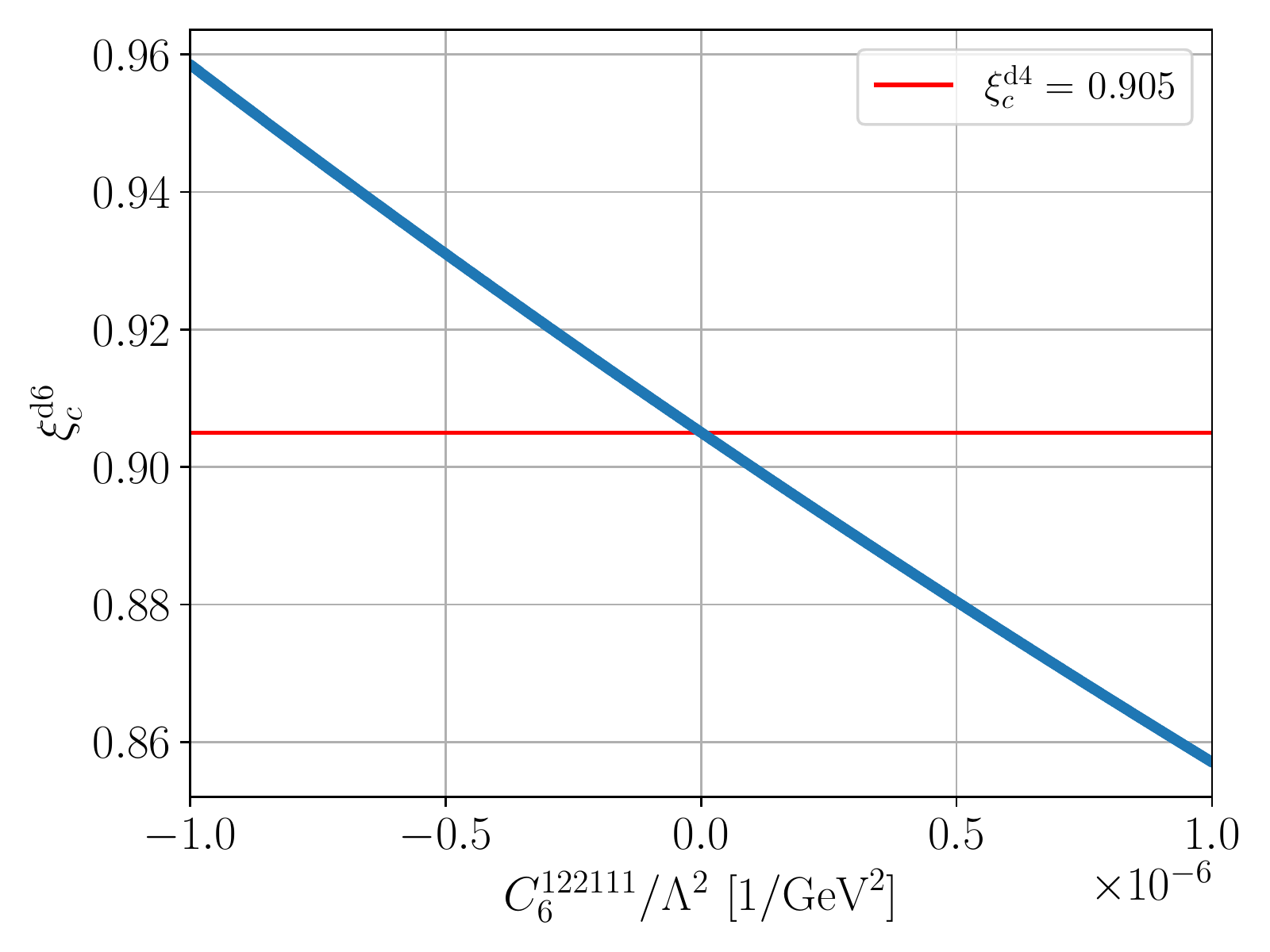}\\
\includegraphics[width=0.43\textwidth]{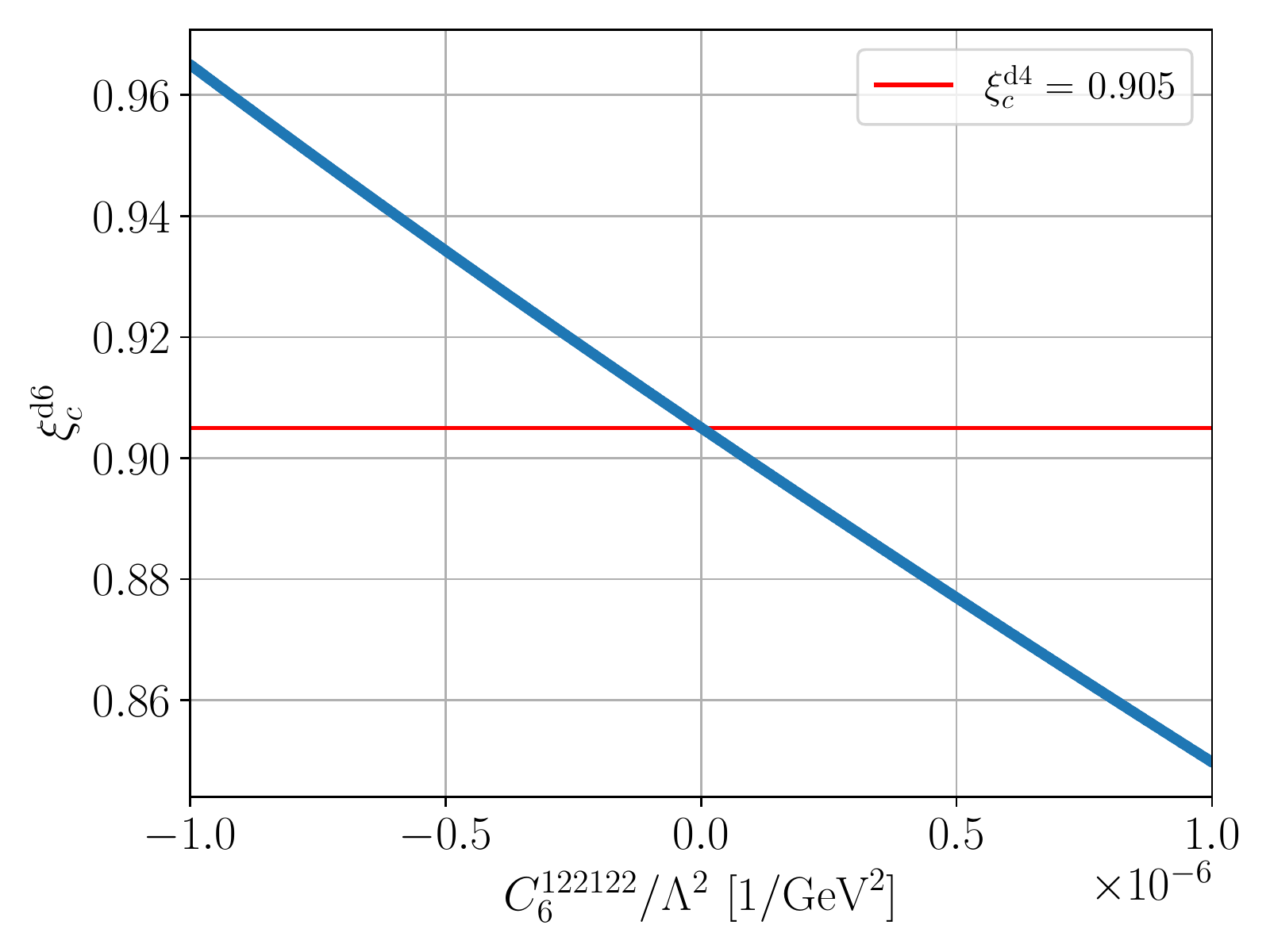}\hspace{0.6cm}
\includegraphics[width=0.43\textwidth]{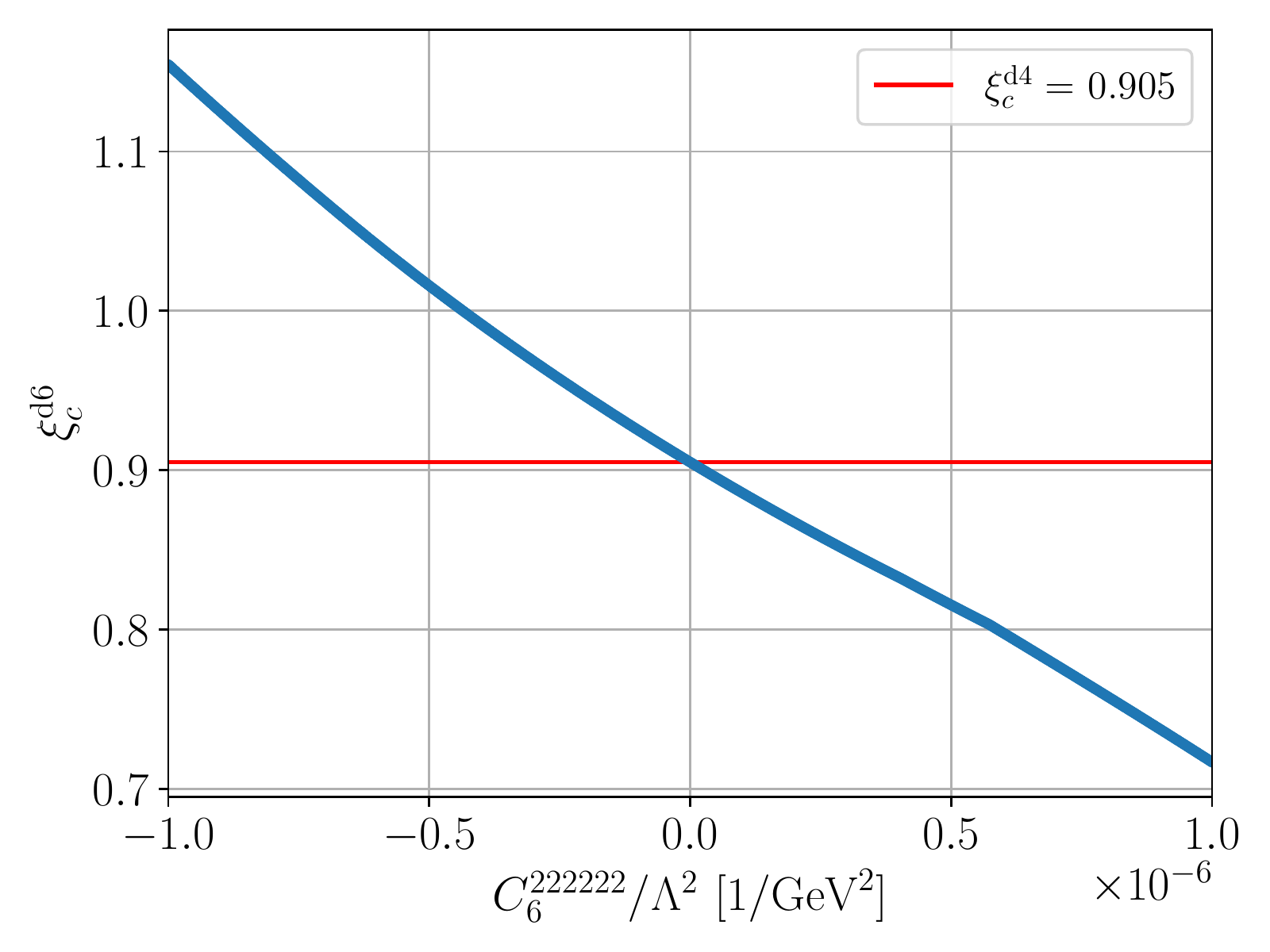}
\caption{Representative behaviour of
  $\xi^{\text{d6}}_c$ for a
  representative parameter point of the 2HDM type II of Tab.~\ref{tab:param}
  with $\xi^{\text{d4}}_c\simeq 0.9$ when the impact of the individual Wilson coefficients is considered. 
  \vspace{0.2cm}
  }
\label{fig:lindep}
\end{figure*}
\begin{table}[b!]
  \begin{center}
      \begin{tabular}{|c|c|c|c|c|c|c|}
          \hline
          $m_{h}$ [GeV] & $m_{H}$ [GeV] & $m_A$ [GeV] & $m_{H^\pm}$ [GeV] & $\tan\beta$ &
          $c_{H VV}$ & $m_{12}^2$ [GeV$^2$] \\ \hline \hline
          125.09 & 681 & 855 & 884 & 1.362 & -0.00459 &
          220945 \\ \hline
      \end{tabular}\\[0.2cm]
      \begin{tabular}{|c|c|c|}
          \hline
          $T_c^\text{d4}$ [GeV] & $v(T_c)^\text{d4}$ [GeV] & $\xi_c^\text{d4}$\\ \hline \hline
          250.55 & 226.76 & 0.91 \\ \hline
      \end{tabular}
  \end{center}
   \caption{\label{tab:param}Input parameters of the benchmark point used for Fig.~\ref{fig:lindep}.}
\end{table}

On the one hand, correlations of masses and couplings are modified away
from the dimension-4 expectation when considering effective
interactions. On the other hand, such deviations are still allowed to
be significant as the LHC has so far only shed limited light on the
structure of the Higgs self-interactions. The choice of input
parameters enables us to directly choose $\alpha$\footnote{In terms of
$c_{HVV}$ in the {\tt ScannerS} scan.}, $\tan\beta$ and the Higgs
boson masses as relevant input parameters of this study. This has the
benefit that electroweak precision constraints and Higgs signal
strengths are largely unchanged from the dimension-4
result\footnote{We explicitly check that the modified charged Higgs
  contributions do not impact
  the SM-like Higgs decay into $\gamma\gamma$.}; the modifications of the dimension-6 interactions of Sec.~\ref{sec:2hdmeft} are then primarily visible in modified Higgs self-interactions ({\it{e.g.}} Higgs pair production). Consistency of single Higgs collider physics observables with the SM follows closely the usual type I or II paradigms~(see \cite{Atkinson:2021eox,Abouabid:2021yvw,Atkinson:2022pcn} for recent analyses).

We first turn to the impact of individual $\Phi^6$ operators of
Tab.~\ref{tab:phi6op} and how they can contribute to a first-order
phase transition.\footnote{As stated above, for all results that we
    present we take parameter scenarios where the lighter of the two
    CP-even Higgs bosons, $h$, is the SM-like one.} Reflecting the fact that these additional
interactions should be small compared to the dimension-4 theory, in first instance, we
consider points that show a relatively strong phase transition
$\xi^{\text{d4}}_c \simeq 0.9$ with the measurements as
reflected in {\tt{HiggsBounds}}, {\tt{HiggsSignals}}, via
{\tt{ScannerS}}. 
As can be seen in Fig.~\ref{fig:lindep}, the effect of additional
contributions to the potential creates to good approximation a linear
dependence $\sim C_6^i$, which demonstrates the robustness of the
approach, {\it{i.e.}} the inclusion of non-linear parts in the $\Lambda^{-1}$
expansion via the Debye masses and the {\tt{BSMPT}} approach is
numerically insignificant. Furthermore, Fig.~\ref{fig:lindep} clearly
shows that an SFOEWPT can be achieved in the 2HDM type II when
considering new effective contributions to the Higgs potential at
perturbative Wilson coefficient sizes in agreement with current
experimental constraints.

The arguably most interesting question then becomes whether the
presence of such operators has collider-relevant implications. The
latter come in a range of different guises; the phenomenology of the
extra Higgs bosons is dominated by top quark final
states, in particular when we are relatively close to the
alignment limit which is favoured by current collider
observations~\cite{Atkinson:2022pcn}. It is well-known that these are
subject to large interference effects that can render the narrow-width
approximation unreliable and could even lead to a vanishing direct
sensitivity for the naively best-motivated LHC
signatures~\cite{Basler:2019nas} (see
also~\cite{Gaemers:1984sj,Bernreuther:1998qv,Dicus:1994bm,Jung:2015gta,Frederix:2007gi,Carena:2016npr,Hespel:2016qaf,Djouadi:2019cbm}). The LHC
experiments include these interference effects to their 2HDM searches,
{\it{e.g.}}~\cite{ATLAS:2017snw,CMS:2019pzc}. In $gg\to H \to t\bar t$ the
exotic Higgs width is the crucial parameter\footnote{Theoretically
  this is apparent from the requirement to evaluate the signal
  component of the process at the complex Higgs
  pole~\cite{Passarino:2010qk,Goria:2011wa} to guarantee
  gauge-independence as a consequence of the Nielsen
  identities~\cite{Nielsen:1975fs}.} and it is therefore worthwhile to
understand the correlation of $\xi_c$ with the Higgs width feeding
into $H \to t\bar t$ searches. Along these lines, the potential lineshape
analysis of a future discovery $H\to t \bar t$ could serve as an
indirect measurement of the  
phase transition in the favoured $t\bar t$
channel. To this end, we define the interference cross section between gluon fusion $gg\to H\to t\bar t$ signal
and QCD $gg\to t\bar t$ continuum as
\begin{equation}
\hbox{d}\sigma^{\text{inf}} \sim 2\text{Re}\,\left\{ {\cal{M}}(gg\to H\to t\bar t) {\cal{M}}^\ast(gg \to t\bar t) \right\}
\end{equation}
where ${\cal{M}}$ denotes the amplitude (we have suppressed identical phase space and parton density factors and work to leading order accuracy in the following).

\begin{figure*}[!t]
\subfigure[Modification of $gg\to H \to t\bar t$ and interference effects with continuum $gg\to t\bar t$.]{\includegraphics[width=0.48\textwidth]{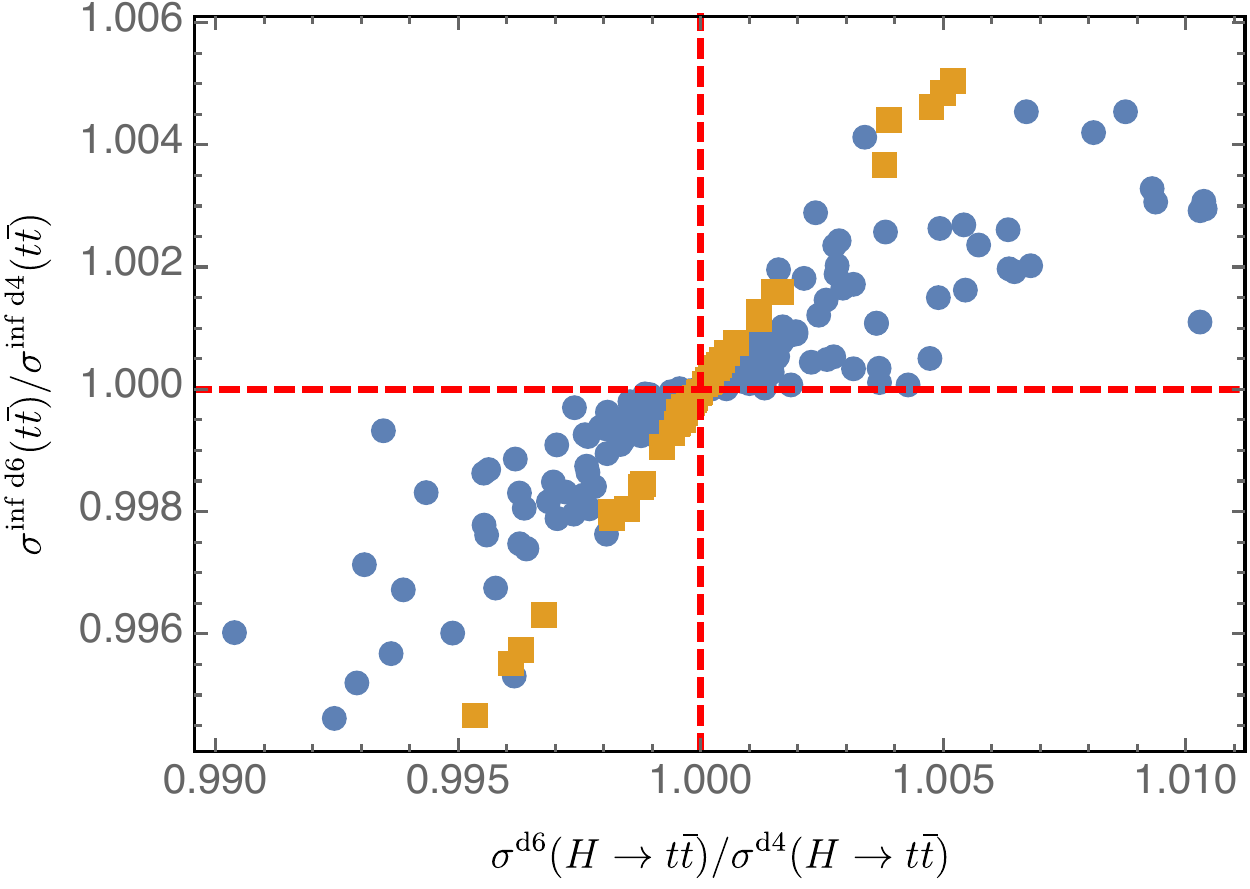}}
\hfill
\subfigure[Modification of Higgs pair production $gg\to hh$ and its
correlation with the resonance contribution $gg\to H\to hh$.]{\includegraphics[width=0.48\textwidth]{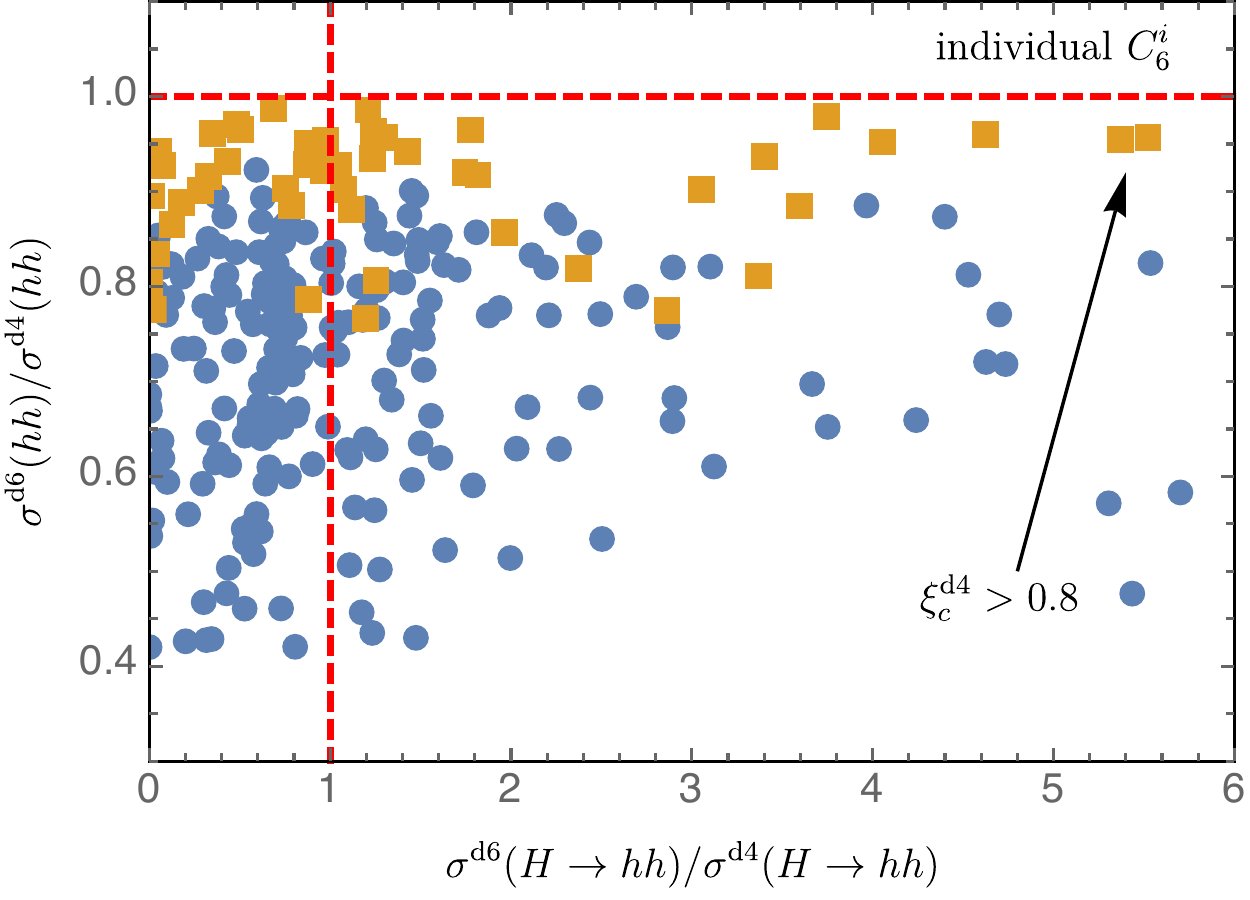}}
\caption{\label{fig:xicxsec} Correlation between EFT-extended cross
  sections and their dimension-4 2HDM counterparts. We scan over parameters that are allowed by the constraints detailed in Sec.~\ref{subsec:scans}, and identify individual Wilson coefficients to achieve $\xi^{\text{d6}}_c=1$. We consider points with $\xi^{\text{d4}}_c>0.3$ and highlight $\xi^{\text{d4}}_c>0.8$ for comparison. For details, see text. 
  }
\end{figure*}

In Fig.~\ref{fig:xicxsec} we show a parameter sample of points in
agreement with the constraints described in Sec.~\ref{subsec:scans}, for
$\xi^{\text{d4}}_c>0.3$ with $\xi^{\text{d6}}_c\simeq 1$ through
choices for single Wilson coefficients (we will study the effect of
combined Wilson coefficients below). We do not distinguish between the individual
Wilson coefficients as the phenomenological outcome is qualitatively
similar. The scan also includes relatively large Wilson coefficient
choices which are necessary to achieve $\xi^{\text{d6}}_c\simeq 1$
starting from  $\xi^{\text{d4}}_c\simeq 0.3$;
for illustration purposes we highlight smaller dimension-6 couplings
resulting from $\xi^{\text{d4}}_c\geq 0.8$ in Fig.~\ref{fig:xicxsec}.
The phenomenological baseline of the $d=4$ points shown in Fig.~\ref{fig:xicxsec} is a top-philic one; $t\bar t$ final states are the preferred 
decay channels of the exotic Higgs bosons with typically
$\text{BR}(H\to t\bar t)\gtrsim 0.8$. The changes that are introduced
by the dimension-6 interactions do not (and to be perturbatively
robust must not) change this behaviour dramatically. In fact, neither
the $t\bar t$ final states, nor their width-sensitive interference
effects show phenomenologically observable modifications,
Fig.~\ref{fig:xicxsec} (a).  There is a trend that reflects the overall
$\xi_c$ behaviour, {\it i.e.} the closer $\xi_c^{\text{d4}}$ gets to unity,
the smaller the $gg\to H \to t\bar t$ modification becomes as a result of a smaller modification of the total $H$ decay width. In any case for the generic top-dominated final states, such per mille level effects are well beyond the sensitivity that can be obtained at hadron colliders.

This leaves multi-Higgs final states as motivated signatures
as shown in Fig.~\ref{fig:xicxsec}~(b). The resonant $H\to hh$
contribution is small as $H\to t\bar t $ is preferred, but there can
be a modification of the resonance signal $gg\to H \to hh$, which is correlated with a modified trilinear $Hhh$ coupling. 
However, the overall $gg\to hh$ rate is decreased. For instance we find 
deviations of 125 GeV Higgs boson pair production of
$\sigma^{\text{d6}}(hh)/\sigma^{\text{d4}}(hh) \simeq 0.4~(0.8)$ for
$\xi^{\text{d4}}_c=0.3~(0.9)$ when sampling individual Wilson
coefficient directions. 
  For large distances $1-\xi^{\text{d4}}_c$ it is clear that the EFT contribution needs to overcome the 2HDM contribution alone, which eventually will put pressure on the dimension-6 EFT assumption, highlighted through non-linear dependencies of $\xi_c^{\text{d6}}(\{C_6^i\})$. The individual Wilson coefficient scans that we have focussed so far remain in their linear regime and hence robust when viewed according to this criterion (this quickly changes for correlated Wilson coefficient choices, see below).

The behaviour of the Higgs pair production cross
sections in Fig.~\ref{fig:xicxsec}~(b) is 
mostly due to the fact that additional potential contributions lead to
an enhancement of the trilinear SM-like Higgs
  self-coupling $\sim \lambda_{hhh}$ at the order of
$\sim50\%$. 
For these coupling
deviations the dominant $gg \to hh$ contribution shows a decreasing
behaviour with a rescaling of the light Higgs self-interaction
$\kappa_\lambda >
1$~\cite{Plehn:1996wb,Baur:2002qd,Baur:2003gpa,Dolan:2012rv,Baglio:2012np,Dolan:2013rja,Frederix:2014hta,LHCHiggsCrossSectionWorkingGroup:2016ypw}. In
the light of existing projections of Higgs boson pair
production~\cite{CMS:2017cwx,Cepeda:2019klc} this can be a manageable albeit
challenging task at the LHC.\footnote{The $t\bar t hh$ final state
  showing an increasing cross section for $\kappa_\lambda>1$ could
  provide additional sensitivity~\cite{Englert:2014uqa}.} The analysis
of the separated resonant $H\to hh$ signal in direct comparison to the $hh$ continuum production can therefore serve as an indirect constraint on
$\xi_c\sim 1$. 
Given that discovery of the $H$ state should become possible in $H\to t\bar t$ first, there is significant 
\begin{wrapfigure}[18]{r}{0.5\textwidth}
\includegraphics[width=.5\textwidth]{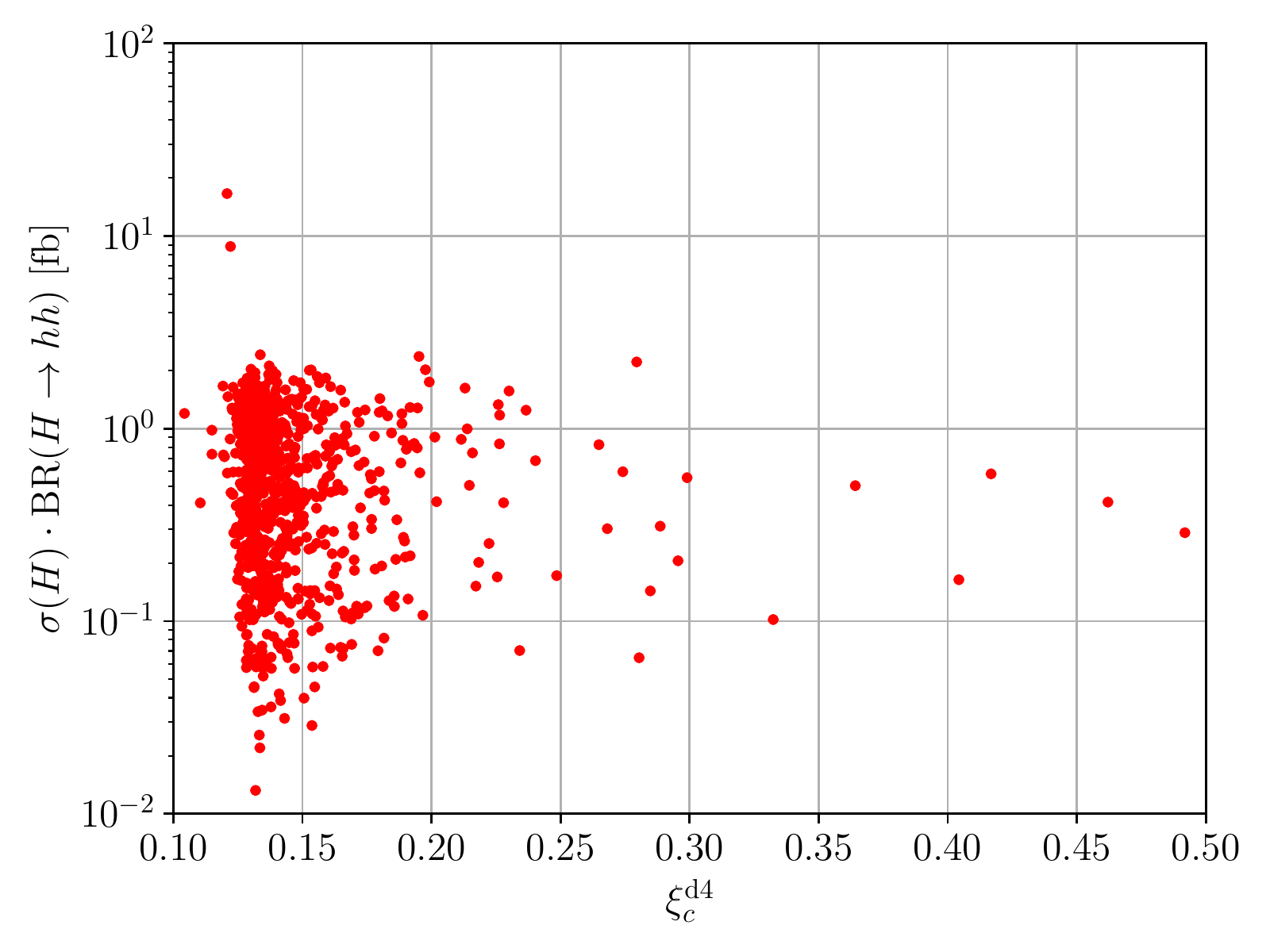}
\caption{\label{fig:largeBR}Correlation of the 2HDM value of $\xi^{\text{d4}}_c$
  and the exotic Higgs production cross section in the double Higgs decay channel.}
\end{wrapfigure}
scope of data-driven methods in separating continuum from on-shell $H$
production.

\begin{figure*}[!t]
\parbox{0.48\textwidth}{\includegraphics[width=.48\textwidth]{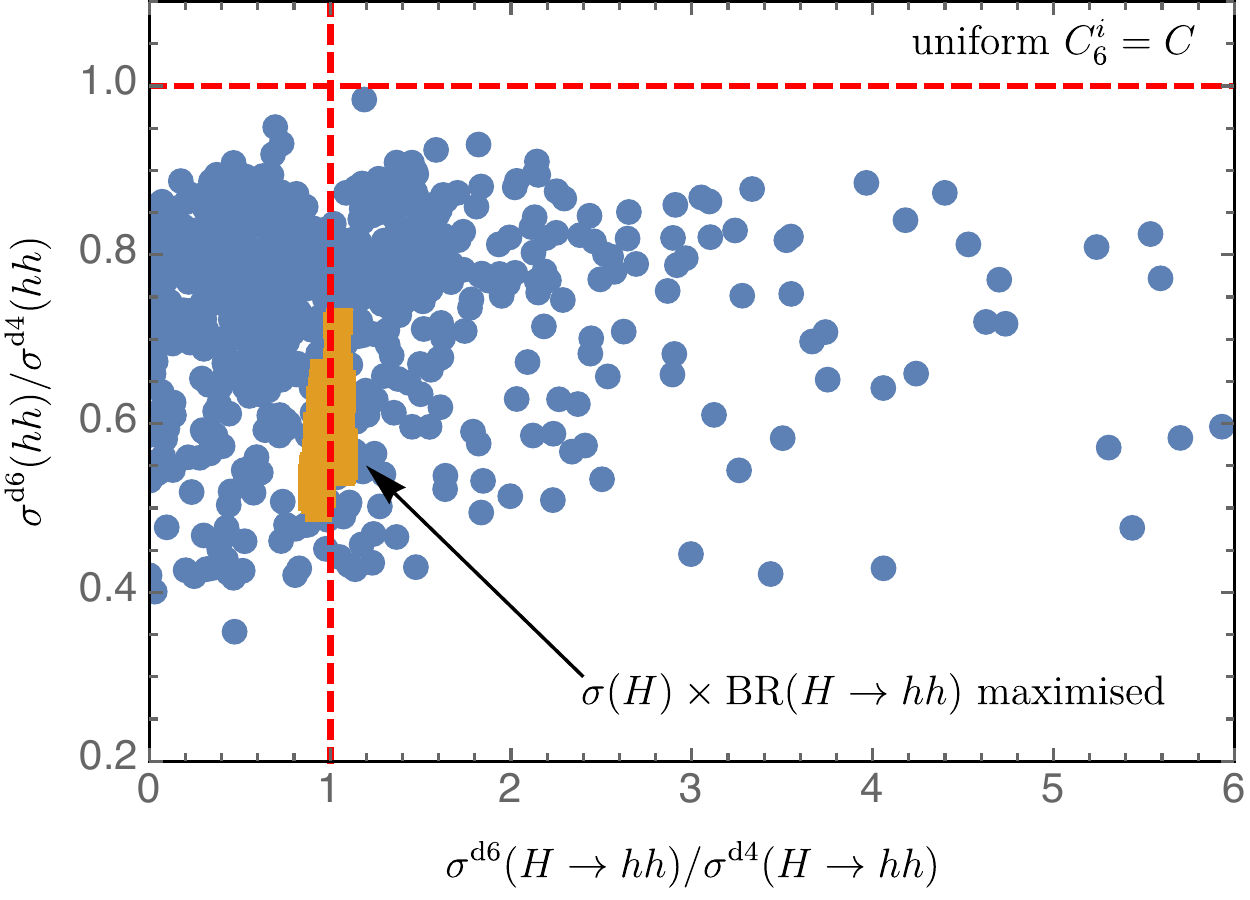}}
\hfill\parbox{0.45\textwidth}{\caption{\label{fig:largeBRxi} Same as Fig.~\ref{fig:xicxsec}(a)  but with uniformly chosen Wilson coefficients $C_6^i=C$ and scanned over $C$ to achieve $\xi^\text{d6}_c\simeq 1$. We highlight the Higgs-philic scan result points of Fig.~\ref{fig:largeBR}, which demonstrates that large potential modifications are required to achieve  $\xi^\text{d6}_c\simeq 1$ in this instance, typically starting from $\xi^\text{d4}_c\simeq 0.15$.}}
\end{figure*}

So far all of our results have been dominated by the general
top-philic nature of the exotic Higgs bosons in the 2HDM. Moving to
larger dimension-4 couplings in the Higgs sector we can furnish
situations where the branching ratio of $H\to hh$ is significant
whilst maintaining reasonable production rates via virtual top
quarks. In such an instance the correlation of on-shell production and
di-Higgs continuum is less statistically limited and therefore
experimentally more feasible. As can be seen in Fig.~\ref{fig:largeBR},
such parameter points in the 2HDM type II are typically characterised
by a larger distance $|1-\xi_c^{\text{d4}}|$. Achieving $\xi_c > 1$ in
a controlled way therefore relies on the interplay of different
effective operators as can indeed be expected in concrete UV scenarios
({\it{e.g.}} in singlet extensions of the scalar
sector~\cite{Profumo:2007wc,Basler:2019iuu,Cho:2021itv,Niemi:2021qvp,Bell:2020gug,Bell:2019mbn}). In
Fig.~\ref{fig:largeBRxi}, we show the results of a scan of uniform
Wilson coefficients $C_6^i=C$ to achieve $\xi^\text{d6}_c\simeq 1$, again for $\xi^{\text{d4}}_c\geq 0.3$. Furthermore, we also include 
the Higgs-philic points of Fig.~\ref{fig:largeBR} to Fig.~\ref{fig:largeBRxi}. These are typically characterised by relatively low $\xi^{\text{d4}}_c$ - the price of Higgs-philic $H$ phenomenology. As can be seen, in general our findings are
similar to the individual Wilson coefficient scan, with large
enhancements possible in the $H\to hh$ rate. As this starts
from a relatively low cross section rate for top-philic $H$ decays,
the largest enhancements $\sigma^{\text{d6}}(H \to h h) /
\sigma^{\text{d4}}(H \to h h)>3$  arise from small $H\to hh$
dimension-4 cross sections. In this instance, a large enhancement is
not directly phenomenologically relevant as the cross section still remains small when including $d=6$ contributions.
Yet, enhancements of factors of $\sim 2.5$ are possible for cross sections in the fb range and we can therefore anticipate some LHC sensitivity here in the $b\bar b b\bar b$~\cite{FerreiradeLima:2014qkf,ATLAS:2022hwc,ATLAS:2020jgy,CMS:2018sxu,CMS:2022cpr} and $b\bar b \tau \tau$ channels~\cite{Dolan:2012rv,ATLAS:2021fet,ATLAS:2020azv,ATLAS:2021tyg,CMS:2017hea,CMS:2017yfv,CMS:2018ipl}.
When turning to points that have a larger $H\to hh$ probability (highlighted in Fig.~\ref{fig:largeBRxi}), the resonance contribution is modified at the 5-10\% level, while the continuum receives a 50\% modification.
We note that for uniform Wilson coefficients squared dimension-6
terms $\sim C_6^i C_6^j /\Lambda^4$ will induce a non-linear behaviour
thus highlighting the importance of a full (matching) calculation to
obtain more realistic estimates. This is particularly relevant for
parameter points with sizeable Higgs-philic branching ratios given in
Fig.~\ref{fig:largeBR}. 
\section{Summary and Conclusions}
\label{sec:conc}
A strong first-order phase transition is a cosmological requirement
within the context of BSM model building. It proves to be non-trivial
as well-motivated SM extensions such as the 2HDM can struggle to
produce a sufficiently large strength in the context of electroweak
baryogenesis. On the one hand, this could mean that baryogenesis
proceeds through mechanisms not associated with the TeV scale.
On the other hand, we can understand and address $|1-\xi^{\text{d4}}_c|$ in terms of additional dynamics that facilitate a strong first-order electroweak phase transition as a minimal modification of the TeV scale.
We have investigated the latter direction in this work for the 2HDM, which remains a strong contender for a more detailed understanding of the electroweak scale. Reverting to effective field theory techniques for the 2HDM, we consider modifications of the Higgs potential at dimension-6 level for the $\mathbb{Z}_2$-symmetric, CP-conserving 2HDM. 
While the 2HDM type II typically falls short of an SFOEWPT, the distance $|1-\xi^{\text{d}4}_c|$ can be overcome by EFT contributions to the Higgs sector.

Additional dimension-6 dynamics that push the 2HDM over the SFOEWPT
finishing line (according to the $\xi_c\simeq 1$ criterion) can then lead to
phenomenological consequences for LHC physics. Interference effects of heavy Higgs
production in the top final state are width-dependent and therefore
sensitive to EFT modifications. The overall effect, however, for the
top-philic final states currently preferred by experimental data
through the alignment limit renders these effects too small to be
measurable at the LHC. 

Higgs pair production is an important tool for fingerprinting an
SFOEWPT, and the distance $|1-\xi^{\text{d}4}_c|$ is directly
correlated with expected Higgs pair production deviation. Current
extrapolations of Higgs pair production to the 3/ab high-luminosity
frontier indicate that the LHC should 
become
sensitive enough to partially explore this region (see also the
recent~\cite{Arco:2022xum}), potentially assisted by discoveries
in the $t\bar t$ channels. For more strongly-coupled Higgs
interactions of the renormalisable 2HDM, the $H\to hh $ signature is
more motivated as a signature for 2HDM discovery. In such an instance,
the 2HDM type II is not capable of producing an SFOEWPT and a
significant modification of the 2HDM Higgs potential is
required. While this stretches the reliability of the dimension-6
approximation, there are phenomenologically relevant implications, predominantly 
the reduction of the $gg\to hh$ rate and modifications of $gg\to H \to hh$.
The LHC is capable of
exploring both phenomenological arenas to some extent and the
discovery of an additional Higgs boson that follows a 2HDM paradigm
could therefore be analysed from an SFOEWPT dimension-6
Higgs-EFT angle. 
\section*{Acknowledgements}
We thank Stephan Huber and Jason Veatch for helpful conversations. This work was funded by a Leverhulme Trust Research Project Grant RPG-2021-031. 
C.E. is supported by the UK Science and Technology Facilities Council (STFC) under grant ST/T000945/1 and the Institute of Particle Physics Phenomenology Associateship Scheme. 
M.M. is supported by the BMBF-Project 05H21VKCCA.


\bibliographystyle{JHEP}
\bibliography{references}

\end{document}